\documentstyle[sprocl_pre,a4,12pt,epsfig]{article}

\def\Journal#1#2#3#4{{#1} {\bf #2}, #3 (#4)}

\def\NPA{{\em Nucl. Phys.} A}

\def\PLB{{\em Phys. Lett.} B}
\def\PRL{\em Phys. Rev. Lett.}
\def\PRA{{\em Phys. Rev.} A}
\def\PRB{{\em Phys. Rev.} B}
\def\PRC{{\em Phys. Rev.} C}

\def\ZPA{{\em Z. Phys.} A}
\def\ZPB{{\em Z. Phys.} B}

\def\ZPD{{\em Z. Phys.} D}

\def\be{\begin{equation}}
\def\ee{\end{equation}}
\def\bea{\begin{eqnarray}}
\def\eea{\end{eqnarray}}

\begin{document}

\title{THE NUCLEAR LIQUID-GAS PHASE TRANSITION:
PRESENT STATUS AND FUTURE PERSPECTIVES
\footnote{\normalsize To appear in the proceedings of the 1st Catania
    Relativistic Ion Studies: Critical Phenomena and Collective Observables,
    Acicastello, May 27-31, 1996.}
}

\author{J. POCHODZALLA}

\address{Max-Planck Institut f\"ur Kernphysik,Saupfercheckweg 1,\\ 
69177 Heidelberg, Germany}

\author{G.~IMME', V.~MADDALENA, C. NOCIFORO, G.~RACITI, 
G.~RICCOBENE,  F.P.~ROMANO, A.~SAIJA, C.~SFIENTI, G.~VERDE }

\address{Dipartimento di Fisica - Universit\'{a} di Catania and
I.N.F.N.- Laboratorio Nazionale del Sud and
Sezione di Catania\\
I-95123 Catania,
Italy}

\author{M.~BEGEMANN-BLAICH, S.~FRITZ, C.~GRO\ss\ , U.~KLEINEVO\ss\ ,
V.~LINDENSTRUTH, U.~LYNEN, M.~MAHI, W.F.J.~M\"ULLER, 
B.~OCKER, T.~ODEH, T.~RUBEHN, H.~SANN,
M.~SCHNITTKER, C.~SCHWARZ, V.~SERFLING, W.~TRAUTMANN,
A.~W\"ORNER, E.~ZUDE}

\address{Gesellschaft f\"ur Schwerionenforschung mbH,\\ 
D-64291 Darmstadt, Germany}

\author{T. M\"OHLENKAMP, W. SEIDEL}

\address{Forschungszentrum Rossendorf,\\
D-01314 Dresden, Germany}

\author{W.D. KUNZE, A. SCH\"UTTAUF}

\address{Institut f\"ur Kernphysik,Universit\"at Frankfurt,\\
D-60486 Frankfurt, Germany}

\author{G.J. KUNDE, S. GAFF, H.~XI}

\address{Department of Physics and Astronomy and National Superconducting
Cyclotron Laboratory, Michigan State University,\\
East Lansing, MI 48824, USA}

\author{R. BASSINI, I.IORI, A. MORONI, F. PETRUZZELLI}

\address{I.N.F.N. Sezione di Milano and Department of
Physics - University of
Milano,\\
I-20133 Milano, Italy}

\author{A. TRZCINSKI, B. ZWIEGLINSKI}

\address{Soltan Institute for Nuclear Studies,\\
00-681 Warsaw, Hoza 69, Poland}


\maketitle\abstracts{
More than two decades ago, the van der Waals behavior of the nucleon - nucleon
force inspired the idea of a liquid-gas phase transition in nuclear matter.
Heavy-ion reactions at relativistic energies offer the unique 
possibility for studying this phase transition in a finite, hadronic system.
A general overview of this subject is given emphasizing the 
most recent results on nuclear calorimetry.
}
  
\section{The Essence of Nuclei}\label{sec:1}


\begin{figure}[htb]
\begin{minipage}[t]{0.48\linewidth}
\centerline{\epsfig{file=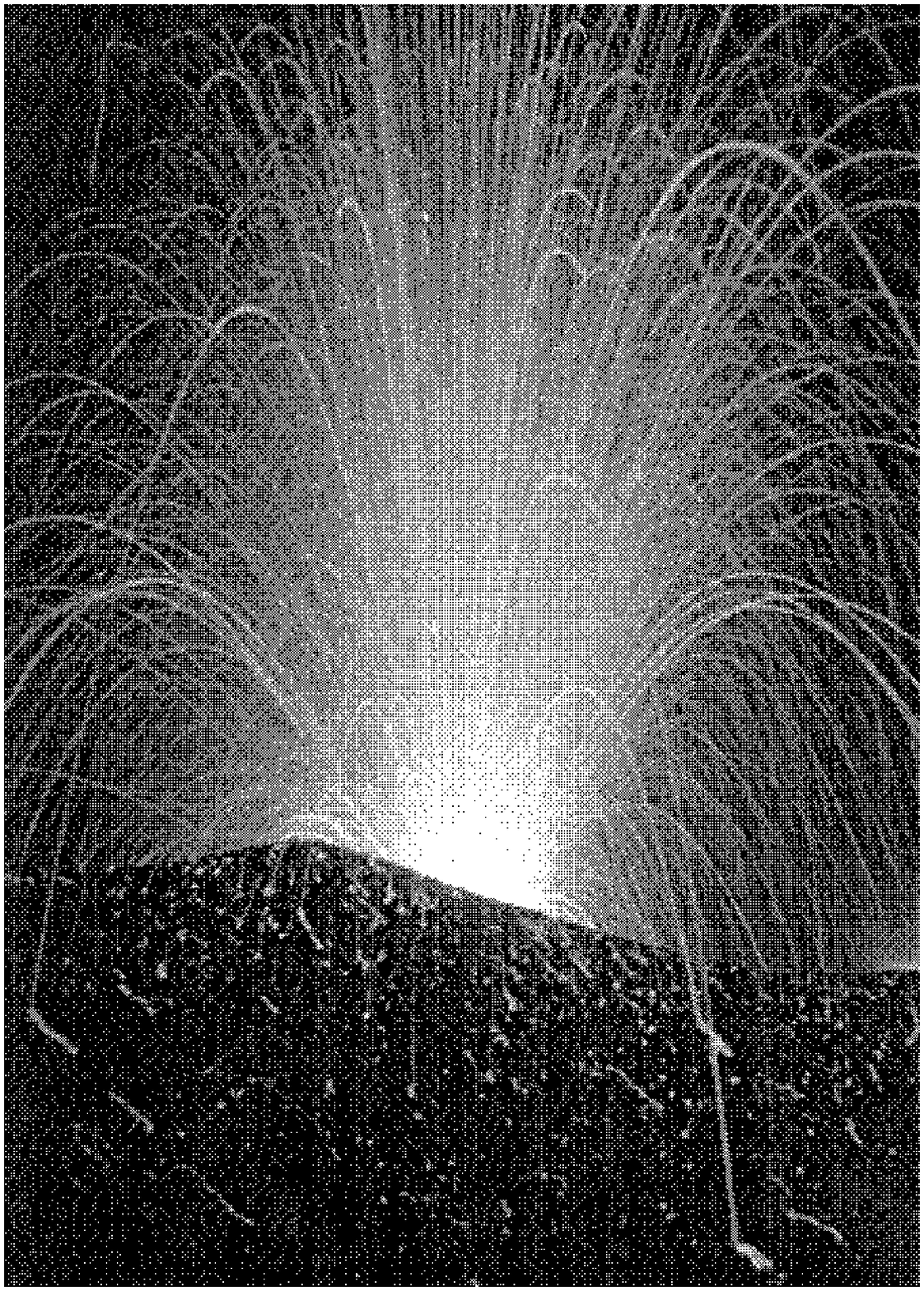,height=8.0cm}}
\end{minipage}
\begin{minipage}[t]{0.48\linewidth}
\centerline{\epsfig{file=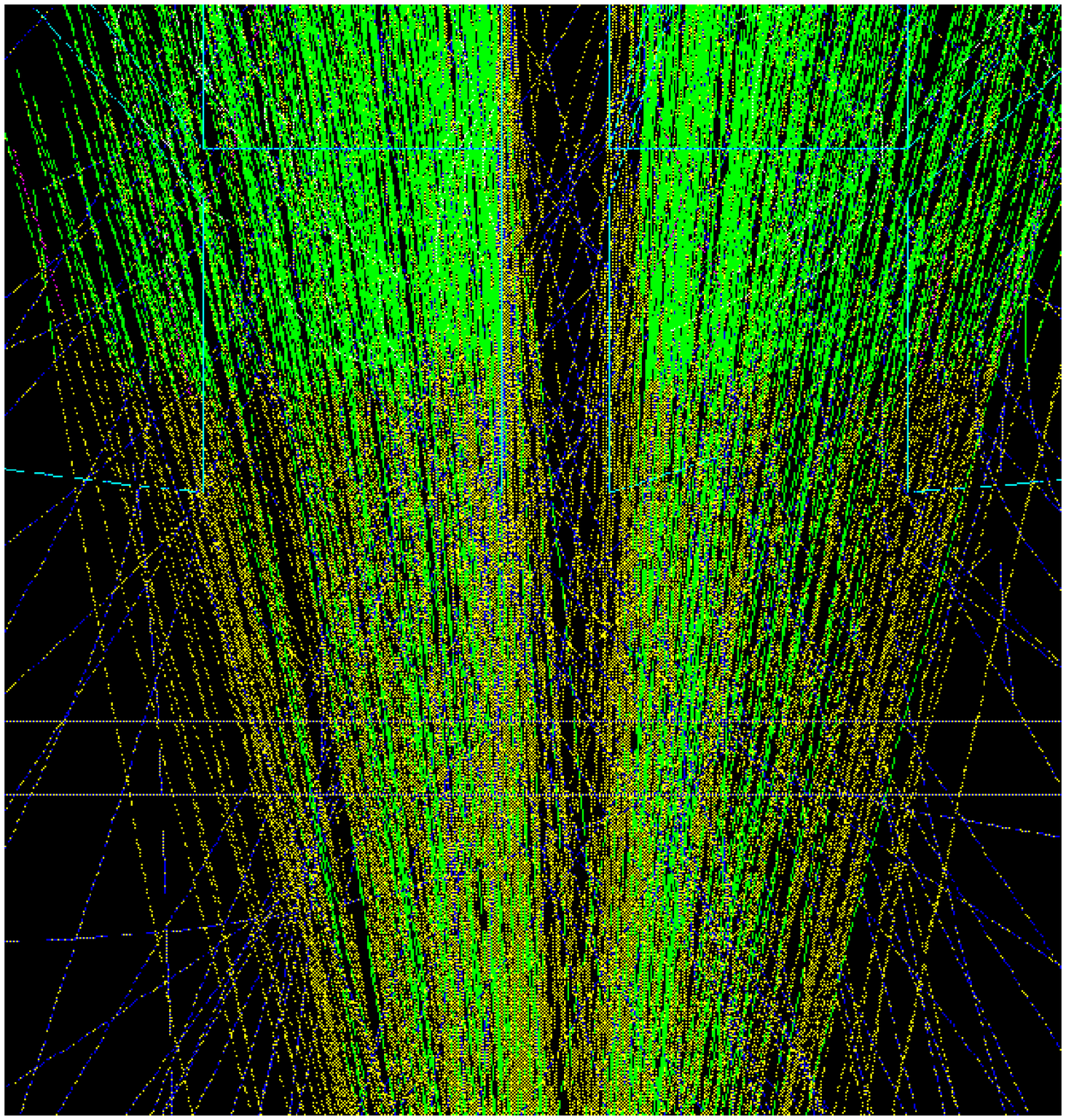,height=8.0cm}}
\end{minipage}

\caption[]{
Left: Empedocles' view of the world.
Right: Expanded view of Pb-Pb collisions at 160 GeV/u recorded by the
NA49 TPC's at the CERN/SPS.}
\label{fig:1}
\end{figure}

Of what are nuclei made? Let us approach this question by asking --
even more general -- of what is the {\it world} made?
If we would have asked Empedocles of Acragas -- 
the Greek philosopher, statesman and poet who 
was born here in Sicily nearly 2500 years ago -- he would have 
answered~\cite{EMP}:
The world is composed of four primal elements
{\it earth, water, air} and {\it fire}. I am sure that the development of these
ideas was inspired by the extraordinary surroundings of his homeland
where everybody could \cite{LUC} -- and still can (Fig.\ref{fig:1}, left) -- 
directly experience the forces of these elements.

Today, of course, we would respond to this question quite differently:
we and the world around us consist out of more than 100
elements, some of which give only a short interlude after their production.
These atoms themselves are made of electrons and nuclei, the latter being
clusters of protons and neutrons. Those are again only combinations of
gluons and quarks and today even quarks can no longer be safely considered
as elementary particles.

Thus, this microscopic view of the world  -- an impressive
illustration of the technical progress made with respect to our
`resolving power' over the last years is shown in the right part of
Fig.\ref{fig:1} -- seems to bring us
close to the final answer of our primary question. But will it provide 
us with a full answer? Is Empedocles' antique view of the world 
already obsolescent? 
Probably not, if we look at the physical world from a more naive point of view. 
For example, a glass of liquid water cooled below 0$^o$C will freeze 
to ice showing the solid character of earth. Water heated above 100$^o$C will
build an airy gas. Finally, at high temperatures atoms disintegrate into 
a plasma of electrons and ions which -- loosely speaking --may be 
considered as the analog to fire.
In that sense the four roots {\it earth, water, air} and {\it fire} 
being representative for {\it solids, liquids, gases} and {\it plasmas}
are omnipresent in our everyday life. 
Today these states of matter are called {\it phases}.

\begin{figure}[t]
\begin{minipage}[t]{0.58\linewidth}
\epsfysize=5.8cm
     \centerline{\epsffile[100 250 500 600]{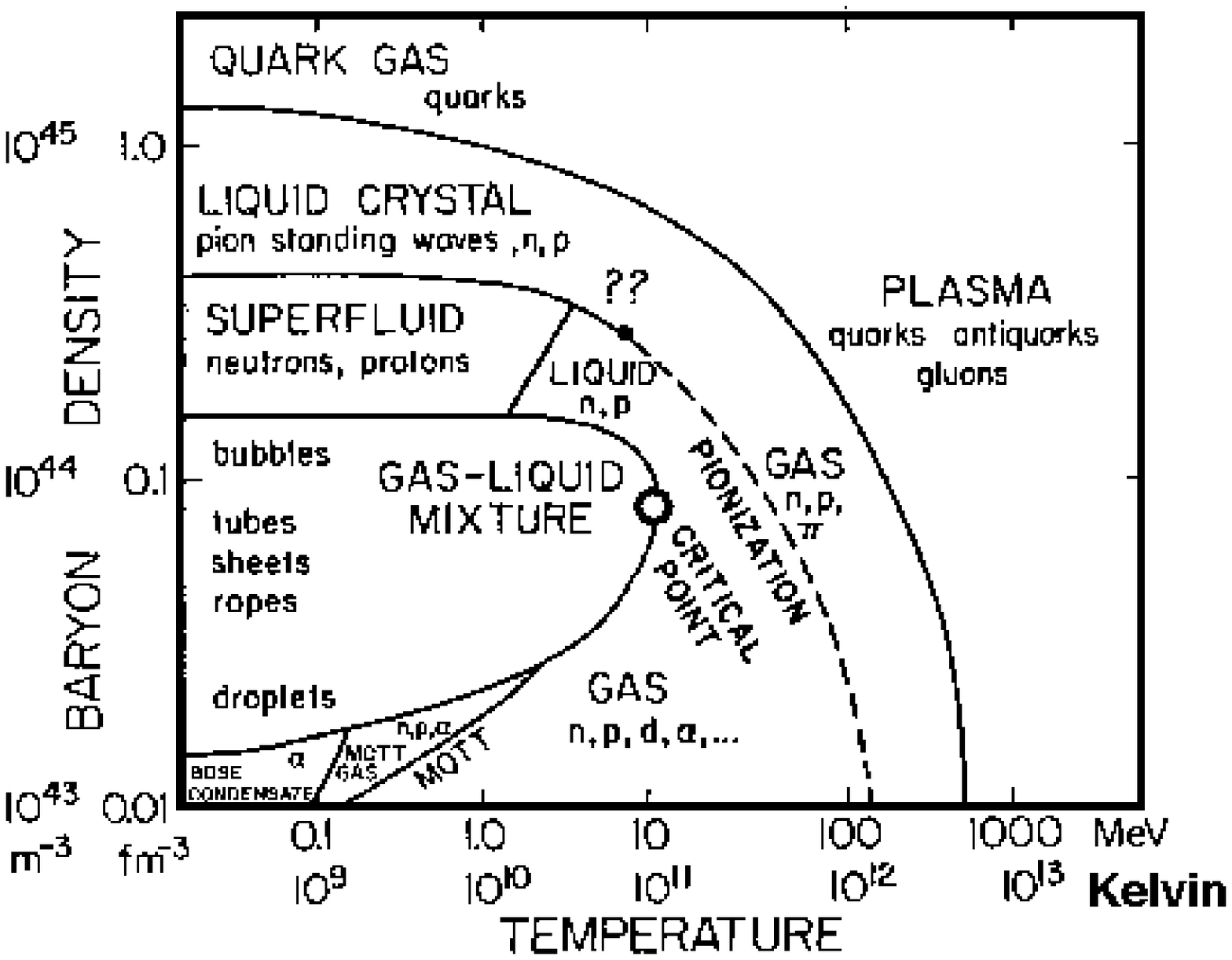}}
\caption [xxxxxx] {
Conceivable landscape of hadronic and nuclear phases \cite{SIE84}.
While most regions of this diagram are
{\it terra incognita}, both, the liquid and the gaseous phase of
nuclear systems are known to exist.
}
\label{fig:2}
\end{minipage}
\hspace{\fill}
\begin{minipage}[t]{0.38\linewidth}
\epsfysize=6.5cm
\centerline{\epsffile [-40 5 430 455]{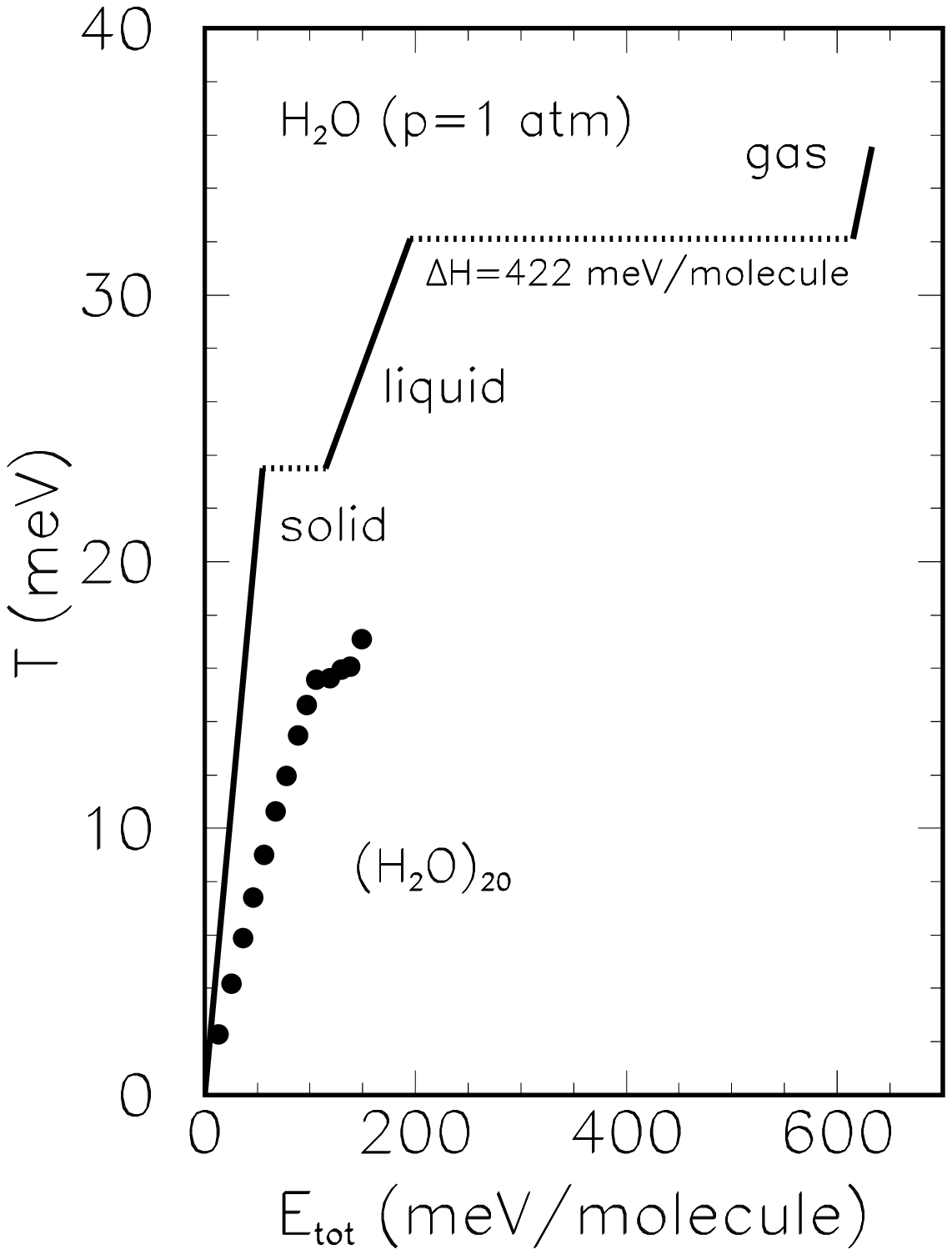}}
\caption[]{
Caloric curve of bulk water at atmospheric pressure (line) and of a water
clusters consisting of 20 H$_2$O molecules predicted \cite{WAL93} by
molecular dynamics calculations (dots).
}
\label{fig:3}
\end{minipage}
\end{figure}

\subsection{The Nuclear Paradigm}\label{sec:1_1}

The prospect of creating a phase of matter resembling that of the
pre-hadronic phase of the early universe or of the core of today's neutron
stars \cite{GLE92}
is one of the prime motivations to study relativistic heavy ion collisions.
Unquestionable, the transition to the quark-gluon plasma represents
the most spectacular example of a phase transition in nuclear matter.
The
complex structure of the hadronic components and the many facets of their
interaction \cite{TAM93}, however,
offer the opportunity to observe in addition several other, exciting nuclear
state transitions (Fig.~\ref{fig:2}, from ref.~\cite{SIE84}).
Already two decades ago, the van der Waals behavior of the nucleon - nucleon
force inspired the idea of a liquid-gas phase transition in nuclear 
matter~\cite{LAM78}$^{\!-\,}$\cite{SIE83}.
What makes this nuclear liquid gas phase transition stand out
from all other conceivable nuclear phase transitions though,
is the fact that both phases, cold nuclear Fermi liquids on the one hand and a
nuclear gas consisting of free nucleons and a few light clusters on the
other hand, are known to exist in nature, and, what may
perhaps be even more important, that both are experimentally accessible.
This unique feature
makes the nuclear liquid-gas phase transition an ideal and relevant
test case for our ability to identify and quantify a phase transition
in a finite hadronic system.

The first observation of a self-similar power law for the fragment mass
distribution in proton
induced reactions was interpreted as an indication for a
critical phenomenon~\cite{FIN82}.
Despite enormous effort during the last decade 
\cite{PAN84}$^{\!-\,}$\cite{CAM88},
the attempts to deduce critical parameters \cite{SIE83} and critical
point exponents~\cite{CAM86}$^{\!-\,}$\cite{ELL94} 
remained elusive~\cite{BAU95}$^{\!-\,}$\cite{MUE96}.
Searching for signals of a nuclear phase transition, we have to cope with
several complications:
Excited nuclei are transient systems which have to be generated in 
nuclear collisions. We are, therefore, facing the difficulty to 
produce isolated
nuclear systems which have reached the highest possible degree 
of equilibration.
Nuclei are composed of a limited number of constituents.
In a finite system, fluctuations are limited.
Singularities of phase transitions get, therefore, rounded and shifted
with respect to their bulk values \cite{IMR80,LAB90}. 
In addition, the long-range Coulomb-repulsion between the constituent protons
introduces additional instabilities \cite{GRO82,LEV85}
which may lead to a further downward shift of the apparent 
`critical'  temperature \cite{SAT89}$^{\!-\,}$\cite{GRO86}.
Since no external fields (e.g. pressure) can be applied 
to excited nuclei in the laboratory \cite{MOR96,POC96}, they may
expand prior to their disassembly \cite{BER83}$^{\!-\,}$\cite{PAP96}.
Eventually, these expanded systems may aggregate into clusters.

Although all these difficulties are inherent in nuclear systems,
they do also apply to other fields where
finite systems are involved \cite{GRO95}.
This is illustrated in Fig.~\ref{fig:3} for the case of water.
The solid line shows the well-known caloric curve of
water at normal pressure of 1 atmosphere.
Comparing this paradigm of first-order phase transitions to the
predicted \cite{WAL93}
caloric curve of a finite
H$_2$O cluster (dots in Fig. \ref{fig:3}) we still recognize the
characteristic `plateau'  signaling  the solid-to-liquid transition. However,
the transition temperature is reduced and distributed over a finite
temperature range.
Nevertheless, this example illustrates that phase transitions of 
rather small clusters
($\sim$10 constituents) are still well defined, distinguishable
\cite{LAB90,WAL93}$^{\!-\,}$\cite{HUE92} 
and -- in case of atomic clusters~\cite{BLU88} --
even detectable, thus nourishing the hope that nuclear systems
produced in energetic heavy ion collisions
may also exhibit sufficiently clear signatures of a phase transition.


\section{The Making of Boiling Nuclei}\label{sec:2}

\begin{figure}[t]
\begin{minipage}[t]{0.48\linewidth}
\epsfysize=6.5cm
     \centerline{\epsffile[-20 50 620 690]{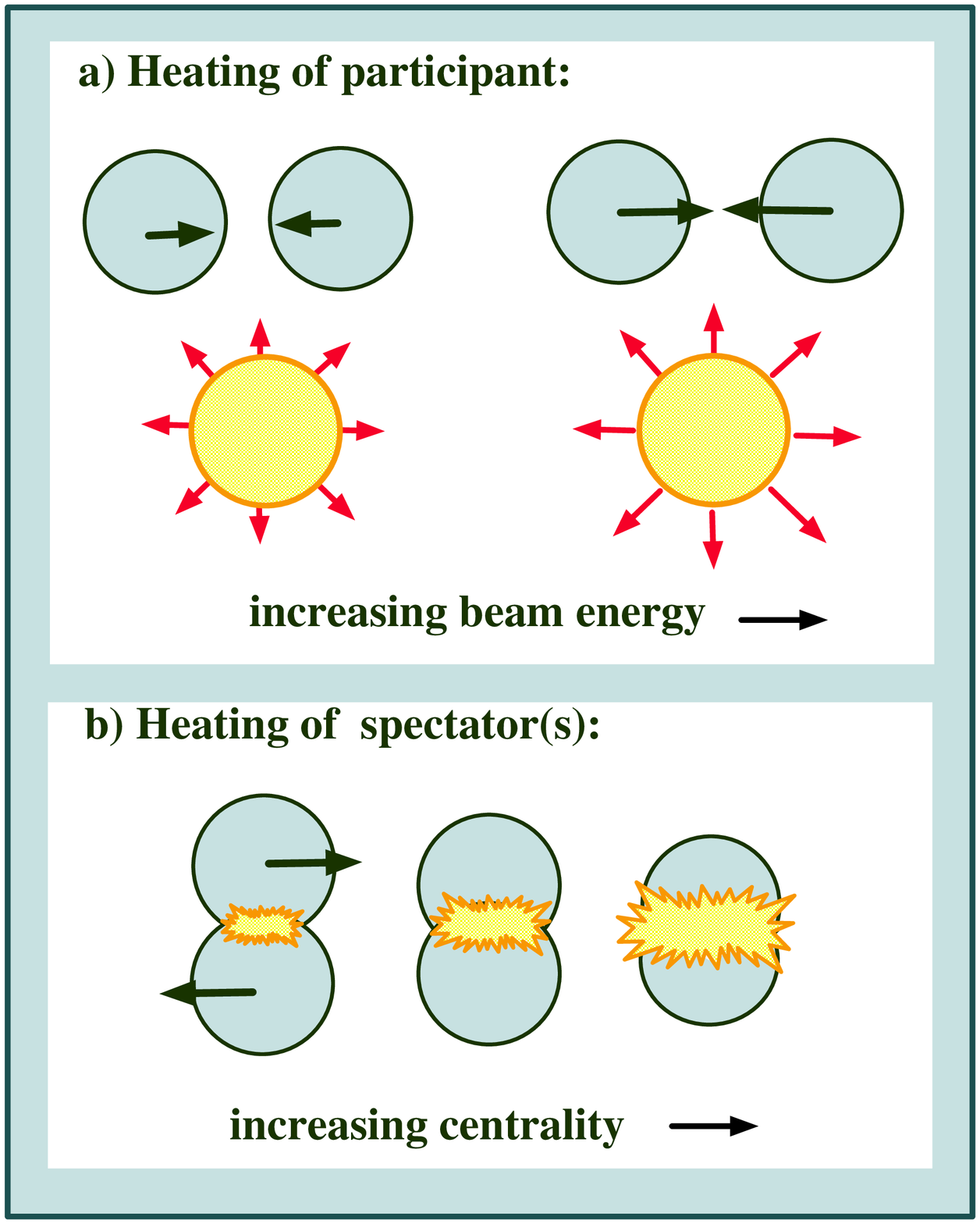}}
\caption [xxxxxx] {
Pictorial view of the two different ways to produce boiling nuclei.
}
\label{fig:4}
\end{minipage}
\hspace{\fill}
\begin{minipage}[t]{0.48\linewidth}
\epsfysize=6.5cm
\centerline{\epsffile [0 20 450 440]{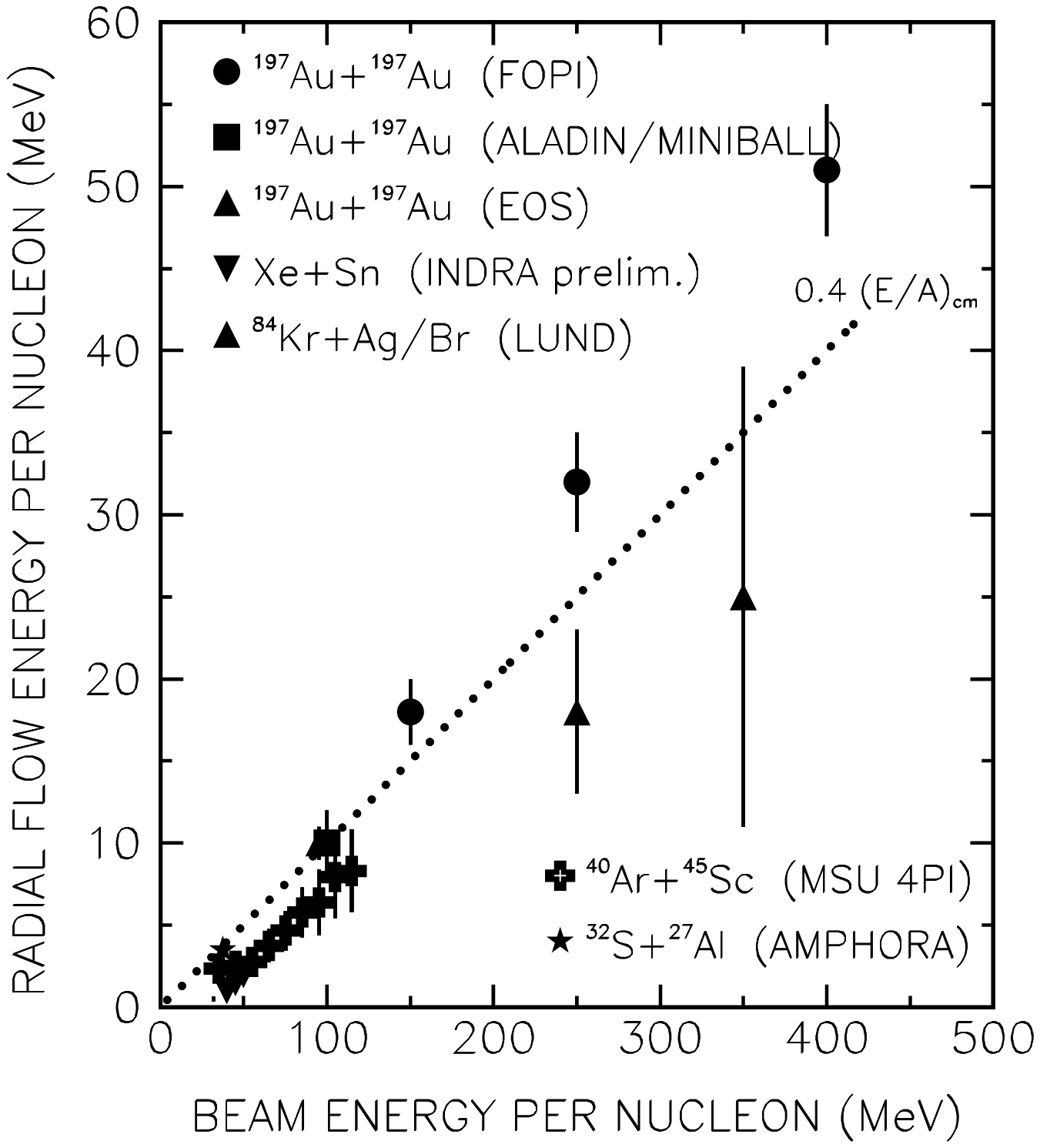}}
\caption[]{
Systematics of radial flow energies in (nearly) symmetric 
nucleus-nucleus collisions as a function of the beam energy per nucleon.}
\label{fig:5}
\end{minipage}
\end{figure}

Heavy-ion reactions at relativistic bombarding energies offer a wide range
of possibilities to produce 
nuclear systems with excitation energies 
around the nuclear binding energy (Fig.~\ref{fig:4}).

In head on collisions between equally heavy nuclei 
the excitation is determined by the incident beam energy. 
The clear advantage of this method is that, for a given 
target-beam combination, systems with nearly
constant mass number can be produced. However, a significant fraction
of the energy is not converted into heat but in collective explosive 
motion, thus introducing an additional degree of freedom.

Spectator nuclei produced in more peripheral collisions do not show this
collective motion in the {\it initial} stage, though
some radial flow may arise during the thermally 
driven expansion \cite{PAP96} and may contribute
to the kinetic energies of the fragments \cite{LAC96} (see also 
chapter \ref{sec:2_3}). 
As it is illustrated in the bottom part of Fig.~\ref{fig:4}, in these
collisions the heating is controlled by the 
size of the fireball and, hence, by the impact parameter.

\subsection{The Little Big-Bang}\label{sec:2_1}

The largest fragment multiplicities measured so far were observed in central
$^{197}$Au on $^{197}$Au collisions at a bombarding energy of 100 MeV per
nucleon \cite{TSA93}.
In these collisions the system of nearly 400 nucleons disintegrates
completely into nuclear fragments and light particles.
An analysis of the kinetic energy spectra in these reactions has revealed a
considerable collective outward motion superimposed on the random motion
of the constituents at the breakup stage (radial flow) which represents 
a significant fraction of the energy available in the
center-of-mass frame \cite{JEO94,HSI94}.
This feature prevails, with monotonously increasing flow values, over the
range of bombarding energies up to 1000 MeV per nucleon
\cite{JEO94}$^{\!-\,}$\cite{PAK96}
(Fig.\ref{fig:5}).

It demonstrates the strong dynamical coupling between
entrance and exit channels in central  collisions of heavy systems.
Indeed, the process of fragment formation is found to be sensitive to the flow
dynamics. The measured relative element yields
are systematically correlated with the magnitude of the observed
radial flow \cite{REI94,KUN95}. Within a coalescence picture, the
suppression of heavier fragments for larger flow values may be caused
by the reduction of the density in momentum space associated
with the collective expansion \cite{KUN95}.

\subsection{The Universality of Spectator Decay}\label{sec:2_2}

\begin{figure}[t]
\begin{minipage}[t]{0.48\linewidth}
\epsfysize=6.0cm
     \centerline{\epsffile[90 230 520 650]{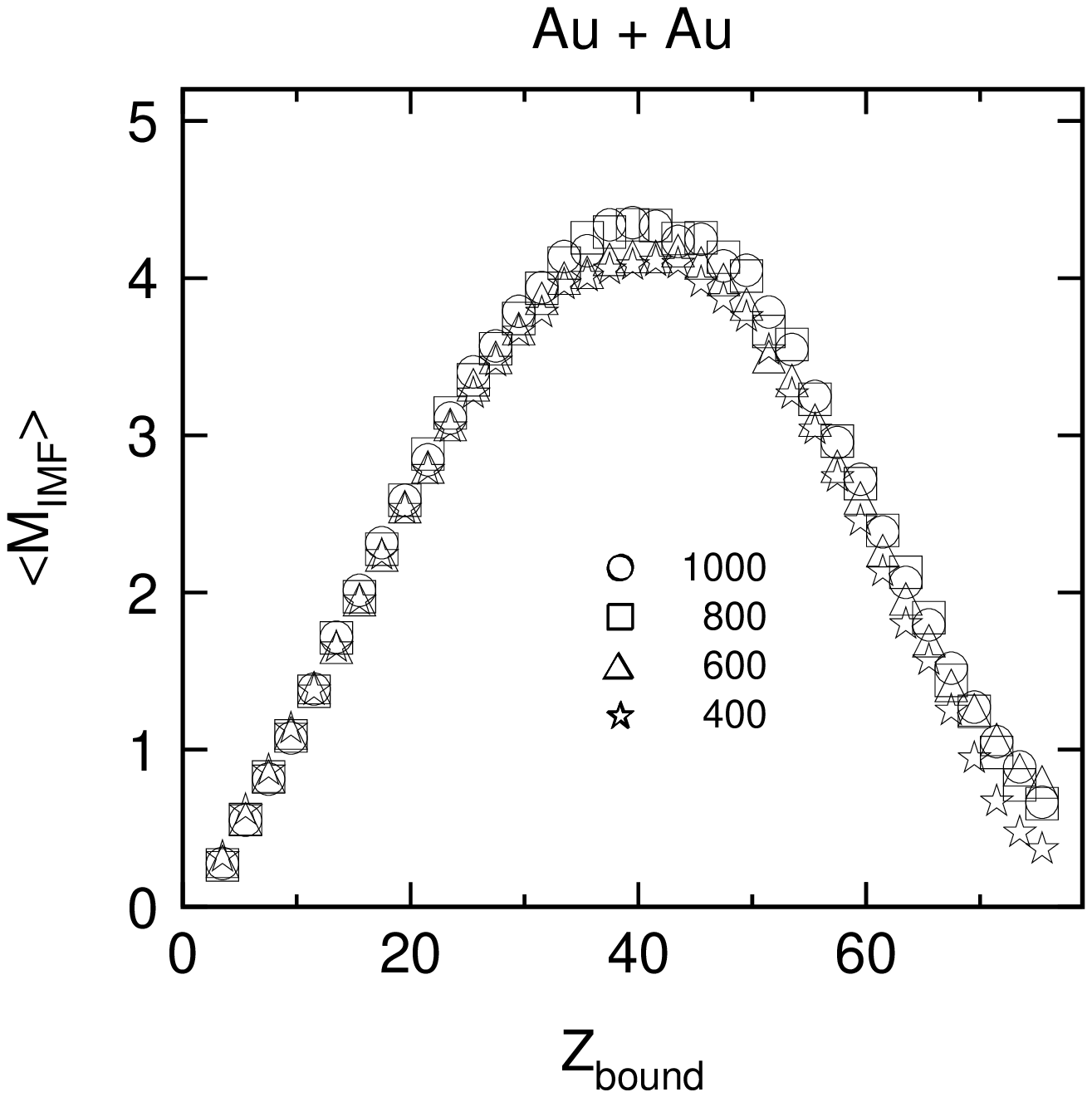}}
\caption [xxxxxx] {
Mean intermediate mass fragment (IMF, 3$\leq Z \leq$ 30)
multiplicity $\langle M_{IMF} \rangle$ as a function of
$Z_{bound}$ for the reaction of $^{197}$Au on $^{197}$Au at $E/A$ =
400, 600, 800, and 1000 MeV
(from ref. \cite{SCH96}).
}
\label{fig:6}
\end{minipage}
\hspace{\fill}
\begin{minipage}[t]{0.48\linewidth}
\epsfysize=6.0cm
\centerline{\epsffile [70 250 550 680]{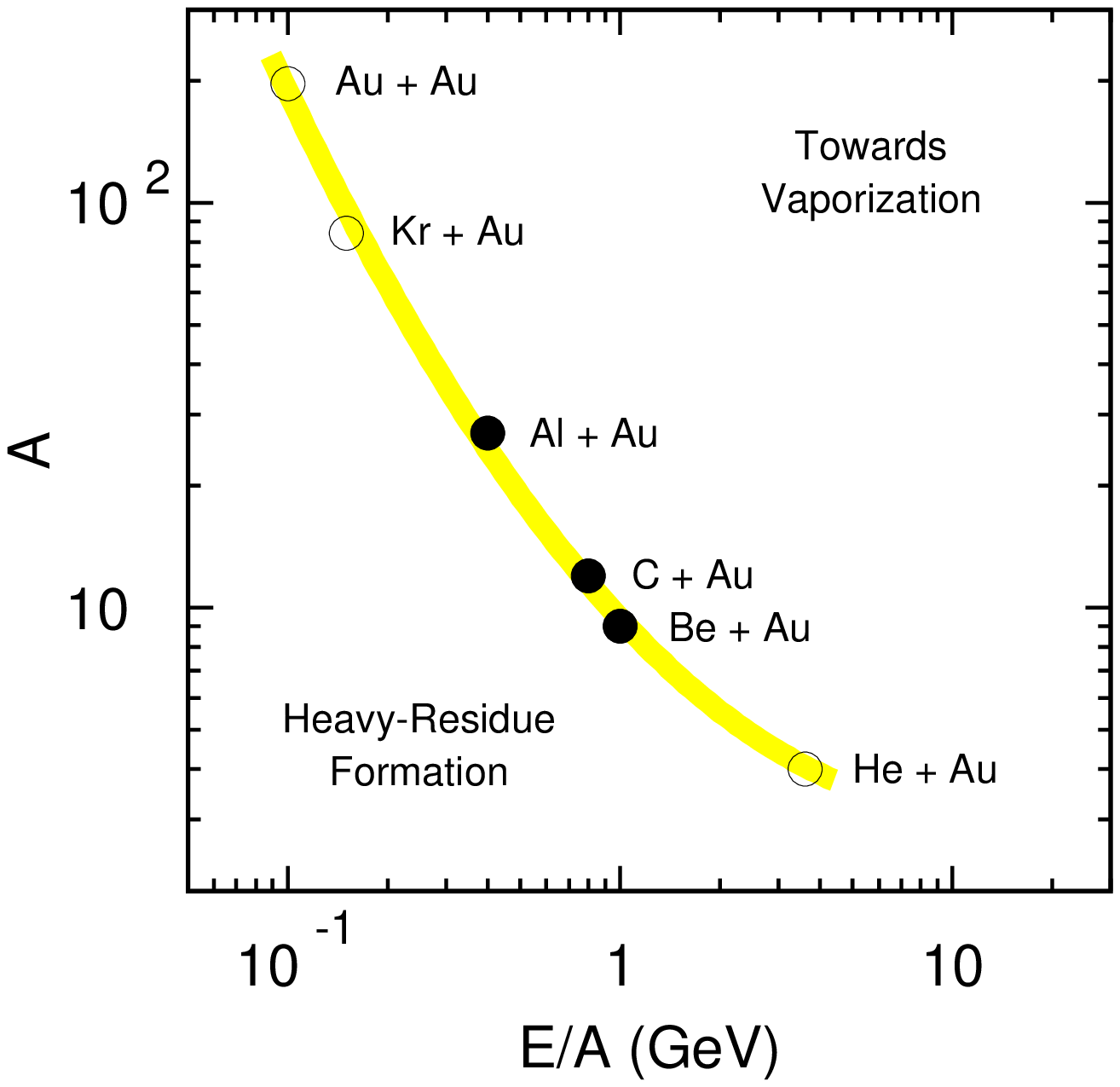}}
\caption[]{
Location of reactions for which maximum fragment production has been
observed in the most violent collisions in a two-dimensional map of projectile
mass versus bombarding energy. Full points denote reactions studied
by the ALADIN collaboration in reverse kinematics, open symbols
refer to work reported in \cite{TSA93,LIP94,PEA94}
(from ref. \cite{SCH96}).
}
\label{fig:7}
\end{minipage}
\end{figure}

While this interplay of dynamical and statistical effects in central
collisions is currently a matter of high interest \cite{BON94}
and remains a challenging subject for future studies,
we will now turn to multifragmentation processes in more peripheral
nucleus-nucleus collisions at high energies.
In contrast to central collisions, no apparent
dependence on the entrance channel is observed in the decay
of spectator nuclei.
The decay patterns were found to be mainly governed by the energy 
transfer to the spectator, as evident from the $Z_{bound}$ scaling of the
fragment charges and their correlations \cite{HUB91,KRE93}.
The quantity $Z_{bound}$ is defined as the sum of the charges of all
product nuclei with $Z \geq$ 2 and is related to the energy deposition.
When plotted as a function of this quantity the fragment multiplicities
and correlations exhibit a universal behaviour. It was first observed
as an invariance with respect to the chosen target in the decays
of $^{197}$Au projectiles at 600 MeV per nucleon \cite{HUB91,OGI91}.

In Fig.~\ref{fig:6} the mean number of intermediate mass fragments 
is shown as a function of $Z_{bound}$
for the reaction of $^{197}$Au on $^{197}$Au at four bombarding energies.
The correlation between the two quantities is seen to be independent of
the projectile energy within the experimental accuracy.
The maximum mean multiplicity of 4 to 4.5 fragments is reached at $Z_{bound}
\approx$ 40.

The observed invariance with respect to the entrance channel is not
restricted to the multiplicity of intermediate-mass fragments but appears
to be a very general feature of the relative asymmetries and other
correlations between the abundance and the atomic numbers of the
fragmentation products \cite{KRE93,SCH96}.
Scaled with the size of the decaying system, the multiplicity of produced
IMF's seems even to be a universal function which is independent of
the mass of the decaying system \cite{SCH96,BEA96}.

The maximum of the IMF production marks the borderline 
between the regime of residue formation and vaporization.
Our data for collisions with rather light targets such as beryllium or
carbon indicate that, in these cases, beam energies considerably above 400
MeV per nucleon are required in order to reach the maximum IMF production
as a dominant process.
Combining the ALADIN results with work reported by other 
groups \cite{TSA93,LIP94,PEA94} we can draw the borderline between the 
`liquid-like' and the `gaseous' regime as shown in
Fig. \ref{fig:7} as a function of the target mass.

\subsection{Energy Deposition}\label{sec:2_3}

The invariant features of the spectator multi-fragment decay are 
consistent with equilibration of the excited systems prior to 
their decay which justifies
interpretations in statistical or thermodynamical terms.
The necessary baseline for such considerations is 
a knowledge of the energy transfer to the excited
primary nuclear system.
On the other hand, the transient nature of finite nuclear systems
makes it difficult to arrive at a unique definition of the decaying
system and its associated excitation energy.
Depending on the observables we consider, it may characterize
different stages of the reaction. In turn, a detailed understanding
of the different type of excitation energies might help to understand
the temporal evolution of the system.

The procedure used to determine the initial mass and excitation energy
of the spectators prior to their decay is very similar to that
proposed by  Campi {\it et al.} \cite{CAM94}.
The basic idea is to determine the invariant mass
(and hence the excitation energy with respect to a nucleus in it's groundstate)
as well as the total charge and mass numbers of the system by
summing up all masses and kinetic energies.

\begin{figure}[t]
\begin{minipage}[t]{0.48\linewidth}
\epsfysize=6.5cm
     \centerline{\epsffile[-40 20 640 710]{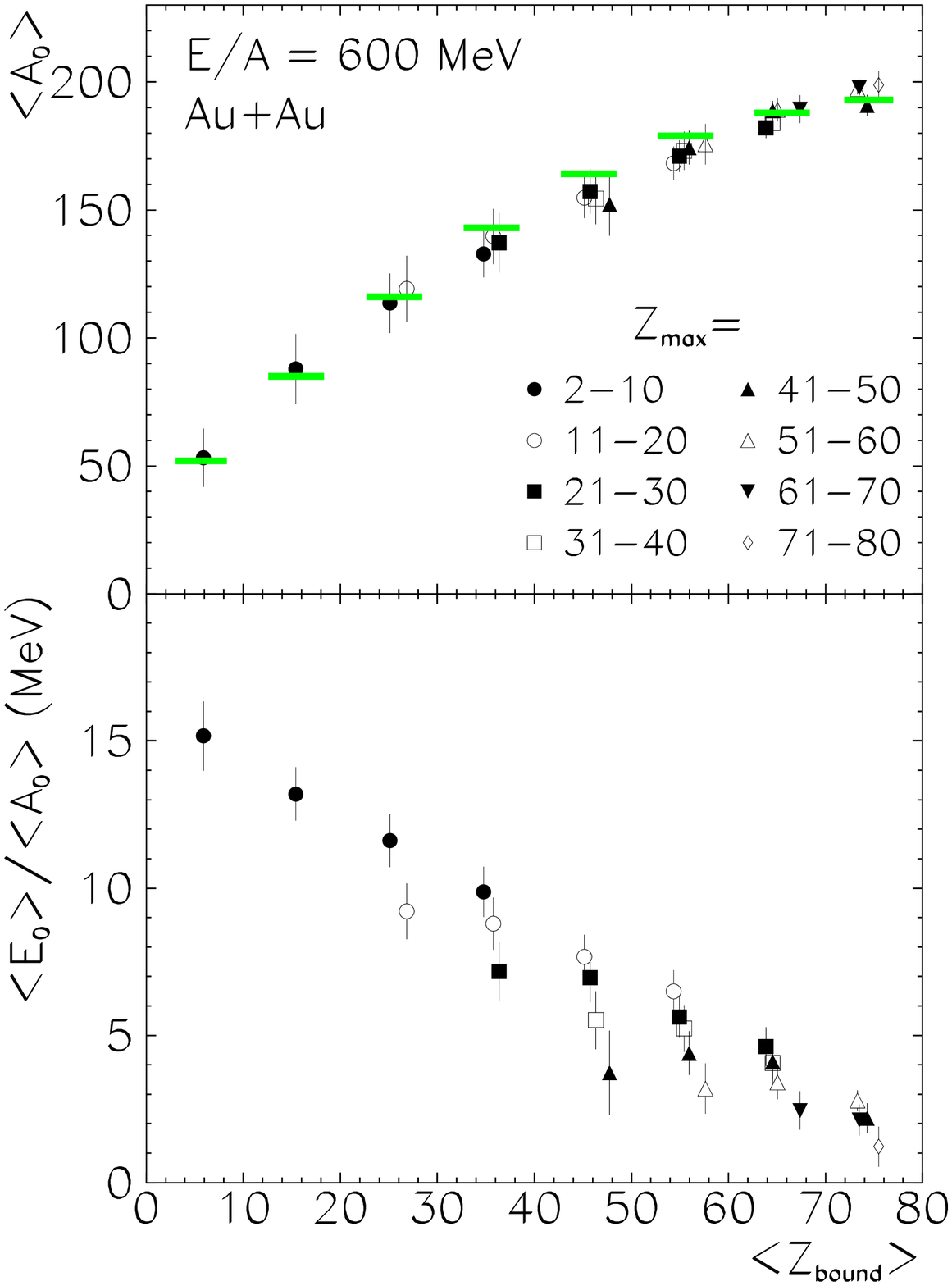}}
\caption [xxxxxx] {
Average prefragment size $\langle A_0\rangle$ and its
excitation energy per nucleon $\langle E_0\rangle$/$\langle A_0\rangle$
as a function of Z$_{bound}$ for different bins in
Z$_{max}$ \cite{POC95}. 
The horizontal bars in the upper part mark the spectator size expected
from a pure participant-spectator geometry. 
}
\label{fig:8}
\end{minipage}
\hspace{\fill}
\begin{minipage}[t]{0.48\linewidth}
\epsfysize=6.5cm
\centerline{\epsffile[-20 10 350 390]{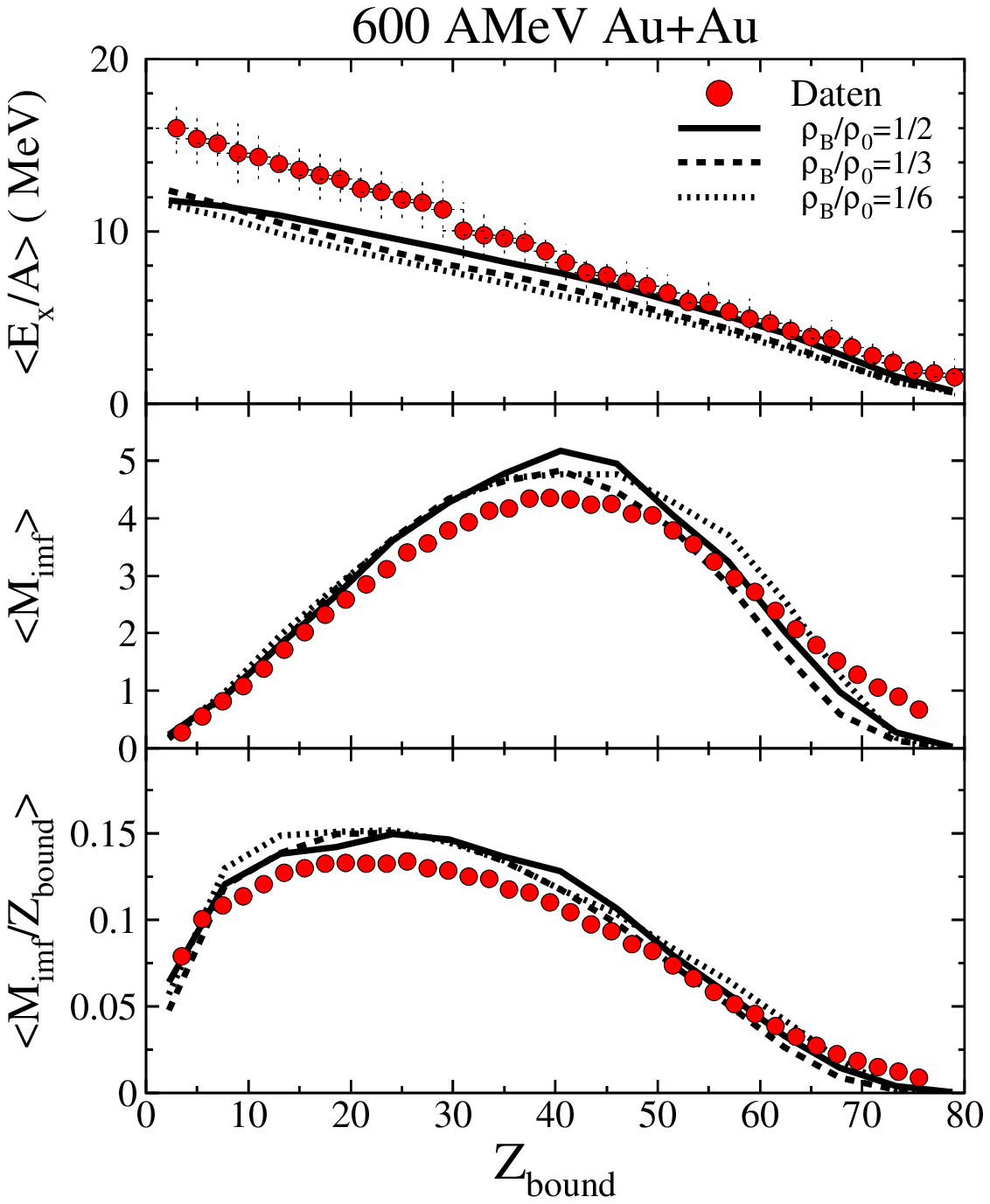}}
\caption[]{
Excitation energy per nucleon (top part), mean IMF multiplicity (middle part)
and normalized IMF multiplicity (lower part) as a function of Z$_{bound}$.
The lines show results of the Moscow statistical multifragmentation 
model \cite{BOT95,KUN96}.
}
\label{fig:9}
\end{minipage}
\end{figure}

The results for the mass $A_0$ and the specific  excitation energy
$E_0/A_0$ are given in Fig.~\ref{fig:8}.
The data points represent the results for individual bins in the
$Z_{max}$-versus-$Z_{bound}$ event representation. The mass $A_0$ decreases
with decreasing $Z_{bound}$ but is, apparently, independent of $Z_{max}$.
The smallest mean spectator mass in the bin of $Z_{bound} \le$ 10 is
$\langle A_0 \rangle \approx$ 50.
Reconstructing the impact parameter from the quantity $Z_{bound}$, one
finds that $\langle A_0 \rangle$ is remarkably well described by
the simple participant-spectator geometry (see horizontal bars in
Fig.~\ref{fig:8}).

The excitation energy $E_0$ appears to be a function of both $Z_{bound}$
and $Z_{max}$; the higher values correspond to the smaller $Z_{max}$
values, i.e. to more complete disintegration of a system of given mass.
The maximum number of fragments, observed at $Z_{bound} \approx$ 40, is
associated with initial excitation energies of  $\langle E_0
\rangle/\langle A_0 \rangle \approx$ 8 MeV.
With decreasing $Z_{bound}$ the deduced excitation energies reach up to
about two times the nuclear binding energy per nucleon.

In the upper part of figure \ref{fig:9} these energy deposits are compared
to predictions of the Moscow statistical multifragmentation model \cite{BOT95}.
In these calculations the model parameters have been adjusted \cite{KUN96}
to describe the observed fragment yields (see for example center 
and lower parts of figure \ref{fig:9}). 
For the range of small $Z_{bound}$ these two energies differ
significantly. Potentially this difference signals the presence 
of (radial) collective motion at the time of breakup 
(see also refs. \cite{LAC96,PAP96}).

\section{Hadronic Thermometer}\label{sec:3}

\begin{figure}[t]
\begin{minipage}[t]{0.50\linewidth}
\epsfysize=8.0cm
\centerline{\epsffile[0 60 620 680]{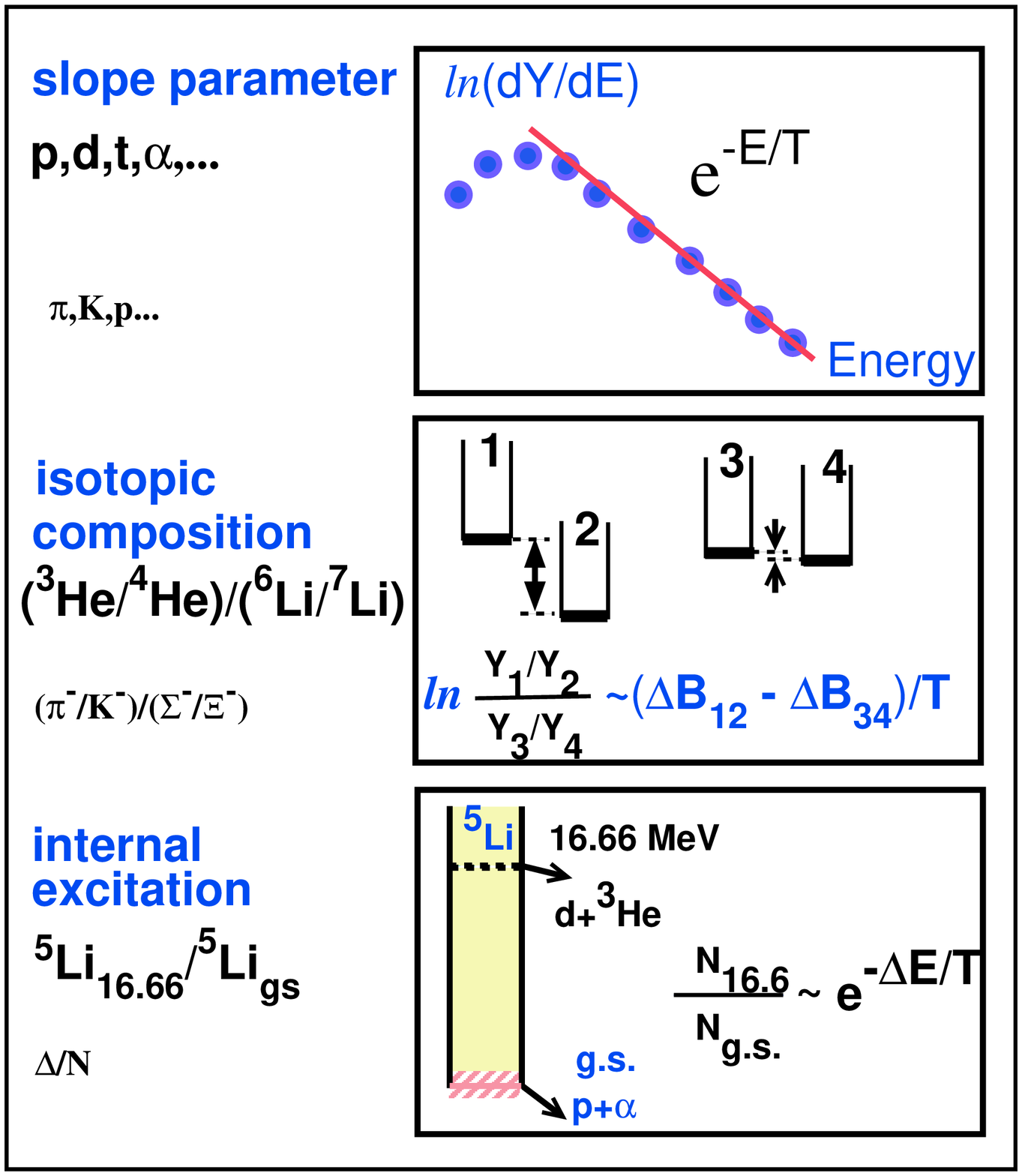}}
\caption[]{
Illustration of different hadronic thermometer:
inverse slope parameter (top part), isotope ratios (center part), and
relative population of states (lower part).
}
\label{fig:10}
\end{minipage}
\hspace{\fill}
\begin{minipage}[t]{0.46\linewidth}
\epsfysize=7.5cm
\centerline{\epsffile [0 0 450 450]{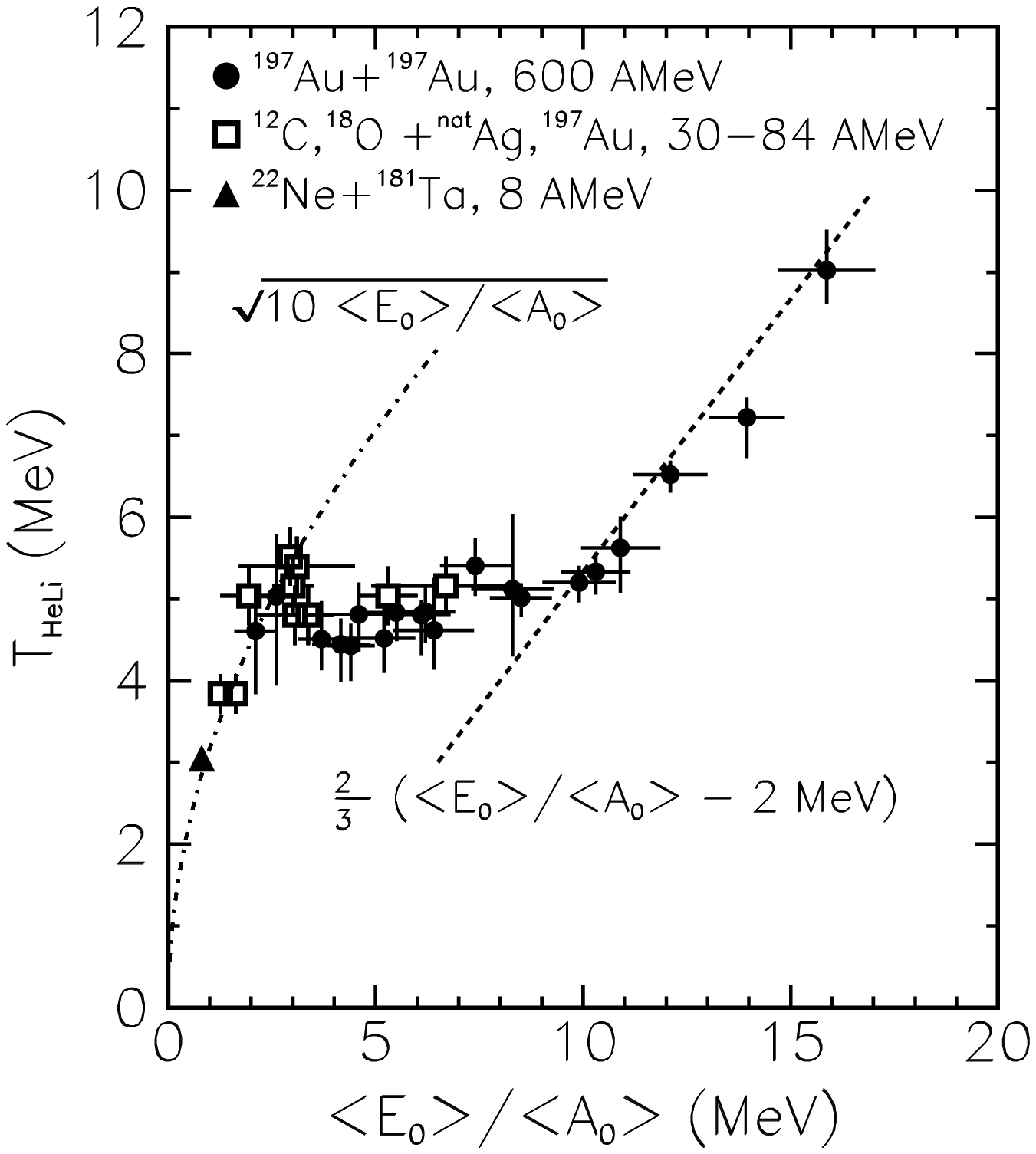}}
\caption[]{
Caloric curve of nuclei determined by the dependence of the
isotope temperature T$_{HeLi}$ on the excitation energy per 
nucleon \cite{POC95}.
}
\label{fig:11}
\end{minipage}
\end{figure}

Less straightforward is the determination of a nuclear temperature.
Nuclei are closed systems without an external heat bath. Consequently,
the temperature of the system cannot be pre-determined
but has to be reconstructed from observable quantities.
For a microcanonical ensemble, the thermodynamic temperature of a
system may be defined in terms of the total-energy
state density.
An experimental determination of the state density and its
energy dependence is, however, hitherto impossible.
Therefore, nuclear temperature determinations take recourse to `simple'
observables of specific degrees-of-freedom 
which constitute  -- at least for some ideal situations and
generally within a canonical treatment -- a good approximation to the true
thermodynamic temperature.

Figure \ref{fig:10} illustrates three different methods to extract
a temperature of a hadronic system.
At low excitation energies, the inverse slope parameters describing the
kinetic energies or transverse mass distributions of the emitted 
particles are a good measure of the temperature. 
In relativistic nucleus-nucleus interactions, however, 
these distributions suffer from possible collective
flow effects and secondary decay processes~\cite{BRO84}$^{\!-\,}$\cite{POC87}. 
While the spectral distributions are indispensable to disentangle thermal 
and collective phenomena, a more direct way to test whether locally thermal
equilibrium is achieved and to determine a temperature 
is to study in detail the particle 
abundance \cite{HOY46,ALB85,HAH88,BRA96,PAN96}. Finally, analysing the
internal population of states of produced fragments, the so called
emission temperature can be deduced. While the latter analysis requires
a more demanding coincidence measurement of the decay products, isotope
temperatures can be extracted from single particle yields.

For the following considerations we will assume a nuclear system 
at low density and in chemical and thermal equilibrium. 
For such a system a measure of the temperature T may be
obtained via the double yield ratio of two isotope pairs, ($Y_1/Y_2$) and
($Y_3/Y_4$), differing by the same  number of neutrons 
and/or protons \cite{ALB85}:

\begin{equation}
   R=\frac{Y_1/Y_2}{Y_3/Y_4} = a \cdot e^{[(B_1-B_2)-(B_3-B_4)]/T}.
    \label{eq:2}
\end{equation}

\noindent
Here, $B_i$ denotes the binding energy of particle species $i$ and the
constant $a$ contains known spins and mass numbers of the  fragments.
Of course, a meaningful temperature scale can only be derived  if the ratio
$R$ is sufficiently sensitive to the temperature of the system and if the
yields of the considered fragments are measurable over a large range of
excitation energy.
A large sensitivity of this thermometer can be achieved if the constant
$b=(B_1-B_2)-(B_3-B_4)$ is larger than the typical temperature to be
measured (see for example ref. \cite{TSA96} for details).

Particularly large values for $b$ are obtained if a $^3He/^4He$ ratio is
involved. Indeed, the large cross section of He fragments as an 
abundant constituent in both the
`liquid' and the `vapor' regime of nuclear systems {\it and} the strong
binding energy of the $\alpha$-particle is the lucky  coincidence which is
the basis of the temperature determination presented hereafter.
In order to acquire  for the second yield ratio also a sufficient
production yield we define in the following a temperature $T_{HeLi,0}$ in
terms of the  yield ratios $^3$He/$^4$He and $^6$Li/$^7$Li
\begin{equation}
  T_{HeLi,0} := 13.33/ln(2.18 \cdot \frac{Y_{6Li}/Y_{7Li}}{Y_{3He}/Y_{4He}}) .
\label{eq:3}
\end{equation}

\noindent
Using the d/t ratio rather than the $^6$Li/$^7$Li ratio,
a similar temperature scale, T${_{HHe,0}}$, may be derived \cite{ALB85}.
Whereas the d/t ratio was not accessible in the first experiment,
T${_{HHe,0}}$ may be more appropriate for central collisions.
In general, by employing four nuclei species which all differ only little in 
proton and neutron numbers, emission from a similar stage of the 
reaction becomes more likely.

Up to this point we considered nuclei in their ground state only and we
ignored the effect of sequential decays during the final stage of the
disassembly process.
In particular, the yield of $\alpha$-particles may be modified by secondary
decays. (Note, however, that the effect of the $\alpha$-feeding is
partially neutralized by the feeding of $^3He$.
Similarly, contributions from $\gamma$-unstable states in $^6$Li and $^7$Li
partially cancel each other.)
In order to test the model dependence of the temperature definition via
Eq.~(\ref{eq:2}) and to investigate the influence of sequential decays and
low lying $\gamma$-unstable states we
analyzed the fragment distributions predicted by several 
decay models \cite{HAH88,KON94,CHA88,GRO86}.
Despite the strong feeding of the light particle yields through secondary
decays these calculations predict an almost linear dependence of
$T_{HeLi,0}$ on the actual temperature T of the system.
However, in order to account pragmatically for a systematic 
underestimation of the temperature by the quantity $T_{HeLi,0}$, 
we define the final isotope temperature via

\begin{equation}
  T_{HeLi} = 1.2 \cdot T_{HeLi,0}.
\label{eq:4}
\end{equation}

\noindent
For consistency reasons all values of T$_{HeLi}$ presented hereafter 
include the factor 1.2.
It is important to realize, though, that this calibration is model dependent
and other decay models might predict different corrections 
(see for example \cite{POC85,TSA96}). 
Also each isotope thermometer will require an individual 
calibration \cite{TSA96,MOE96}.
This model dependence may only be reduced if more data on the population 
of excited states and the fragment distribution become available.

\section{Nuclear Calorimetry}\label{sec:4}

Figure \ref{fig:11} shows the isotope temperature
as a function of the total excitation energy per nucleon \cite{POC95}.
This caloric curve can be divided into three distinctly different sections.
In line with previous studies in the fusion evaporation 
regime~\cite{FAB87}
the rise of T$_{HeLi}$ for excitation energies below 2 MeV per nucleon
is compatible with the low-temperature approximation of a fermionic system.
Within the range of $\langle E_0\rangle /\langle A_0\rangle$ from
3 MeV to 10 MeV an almost constant value for $T_{HeLi}$ of about 4.5-5 MeV is
observed. This plateau may be related to
the finding of rather constant emission temperatures
over a broad range of
incident energies which were deduced from the population of
particle unstable levels in He and Li fragments \cite{KUN91}.
We also note that the mean excitation energy of the
plateau coincides with the limiting excitation energy
for the fusion-evaporation process
of about 4.5-6.4 MeV per nucleon \cite{BOH90}.
Finally, beyond a total excitation energy of 10 MeV per nucleon,
a steady rise of $T_{HeLi}$ with increasing
$\langle E_0\rangle /\langle A_0\rangle$ is seen.

\subsection{Caloric Curve of the Nuclear Fireball}\label{sec:4_1}

\begin{figure}[htb]
\epsfysize=7.0cm
\centerline{\epsffile[60 30 380 350]{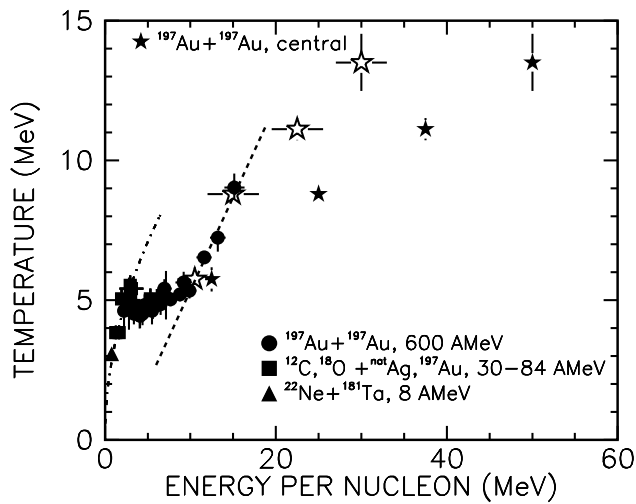}}
\caption [xxxxxx] {
Caloric curve of nuclei determined by the dependence of the
isotope temperature T$_{HeLi}$ on the excitation energy per nucleon.
The stars indicate results for central Au+Au collisions
at 50, 100, 150 and 200 MeV per nucleon incident beam energy.
For the filled stars the energy scale is given by the center-of-mass energy
whereas in case of the open stars the radial flow energy 
(figure \ref{fig:5}) has 
been subtracted.
For orientation the lines given in Fig. \ref{fig:11} are also shown.
}
\label{fig:12}
\end{figure}

While in central collisions between equally heavy nuclei the 
slope parameters and the collective radial 
motion are strongly interlaced, the chemical temperatures deduced from the
isotopic composition reflect a local property
and are expected to be less affected by a radial flow.
Indeed, within the simple coalescence picture outlined in ref. \cite{KUN95}
the double ratio entering into the evaluation of
T$_{HeLi}$ will be modified by not more than 5\% for a typical
ratio between flow and thermal energy of one. 

The filled stars in Fig.~\ref{fig:12} show values for T$_{HeLi}$
for central Au+Au collisions at beam energies of 50, 100, 
150 and 200 MeV per nucleon~\cite{SER96}.
Central reactions were selected by the number of light particles
detected in the forward hemisphere in the center-of-mass~\cite{50MEV}.
Isotope ratios measured close to 90$^o$ in the cm-system were
used to evaluate the isotope temperatures~\cite{RAC96}.
For these data points, the {\it total}
available center-of-mass energy per nucleon has been chosen
as the horizontal axis.
However, as discussed in chapter \ref{sec:2_1}, only part of this energy
is available for heating. For a proper comparison with the caloric curve
determined by the spectator nuclei, one had to determine the
thermal excitation energy {\it at normal density}. A lower limit for this  
energy can be obtained by subtracting the whole measured flow 
from the center-of-mass energy. The corresponding data
points are indicated by the open stars in Fig. \ref{fig:12}. 
Even considering the fact that the flow energy generated during
the expansion from normal nuclear density towards the freeze-out
density should be included in the energy scale, the similarity
between the caloric curves in central and peripheral
collisions is quite impressive and may be viewed as a signal
of common underlying physics.
Of course, a more quantitative understanding of the expansion dynamics
will be required before the question can be answered whether and to what
extent radial flow modifies the properties of the caloric curve.   

\begin{figure}[htb]
\begin{minipage}[t]{0.53\linewidth}
\epsfysize=7.5cm
\centerline{\epsffile [-30 10 490 455]{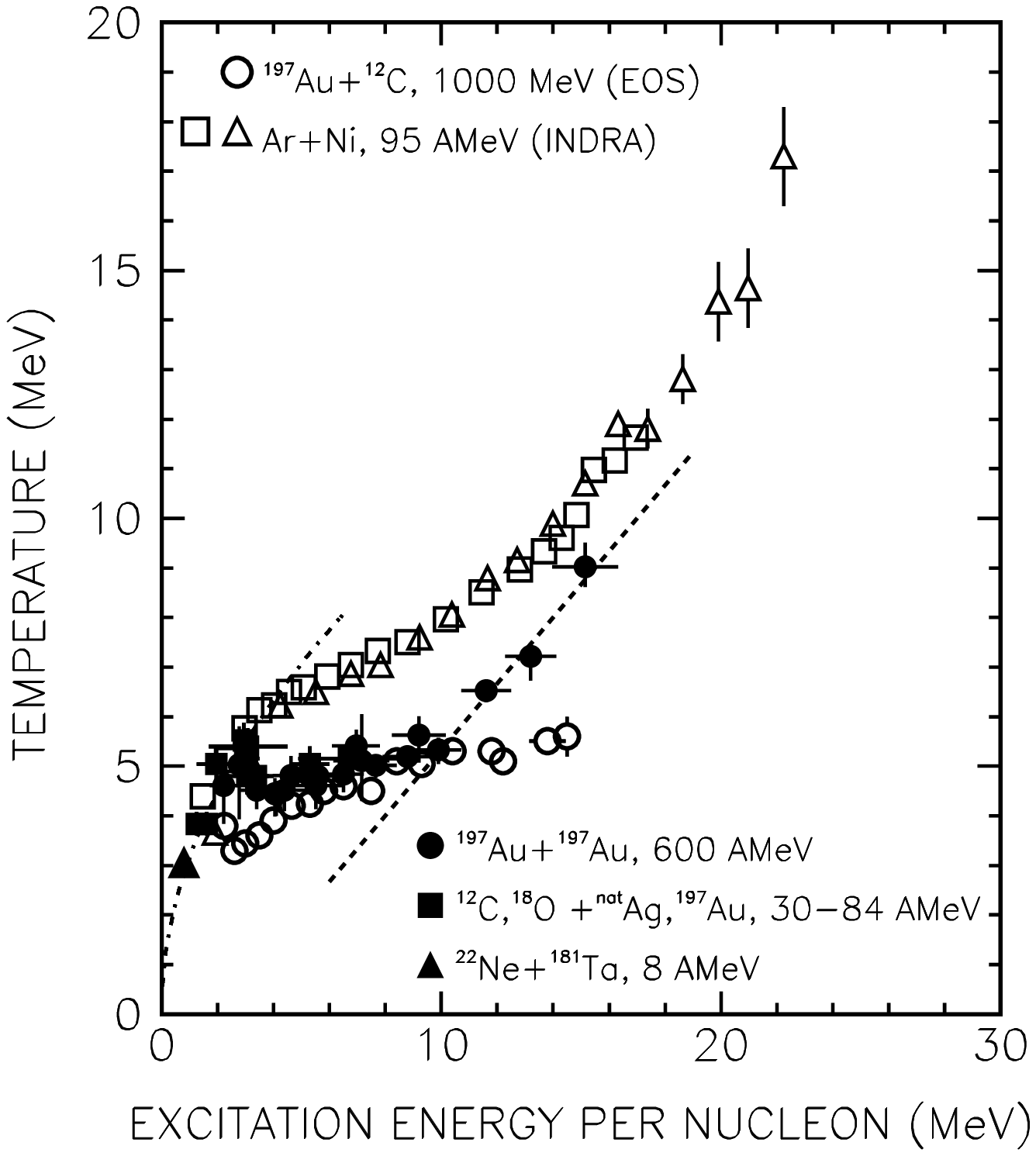}}
\caption [xxxxxx] {
Comparison of ALADIN's caloric curve (solid points) to results obtained by the
EOS collaboration for spectators produced in Au+C reactions at 1000 AMeV
\cite{TIN96,HAU96}
(open circles)
and by the INDRA collaboration for quasi-projectiles
produced in 95 AMeV Ar+Ni reactions \cite{PET96,AUG96} 
(open triangles and squares).
}
\label{fig:13}
\end{minipage}
\hspace{\fill}
\begin{minipage}[t]{0.43\linewidth}
\epsfysize=7.5cm
\centerline{\epsffile [50 190 550 690]{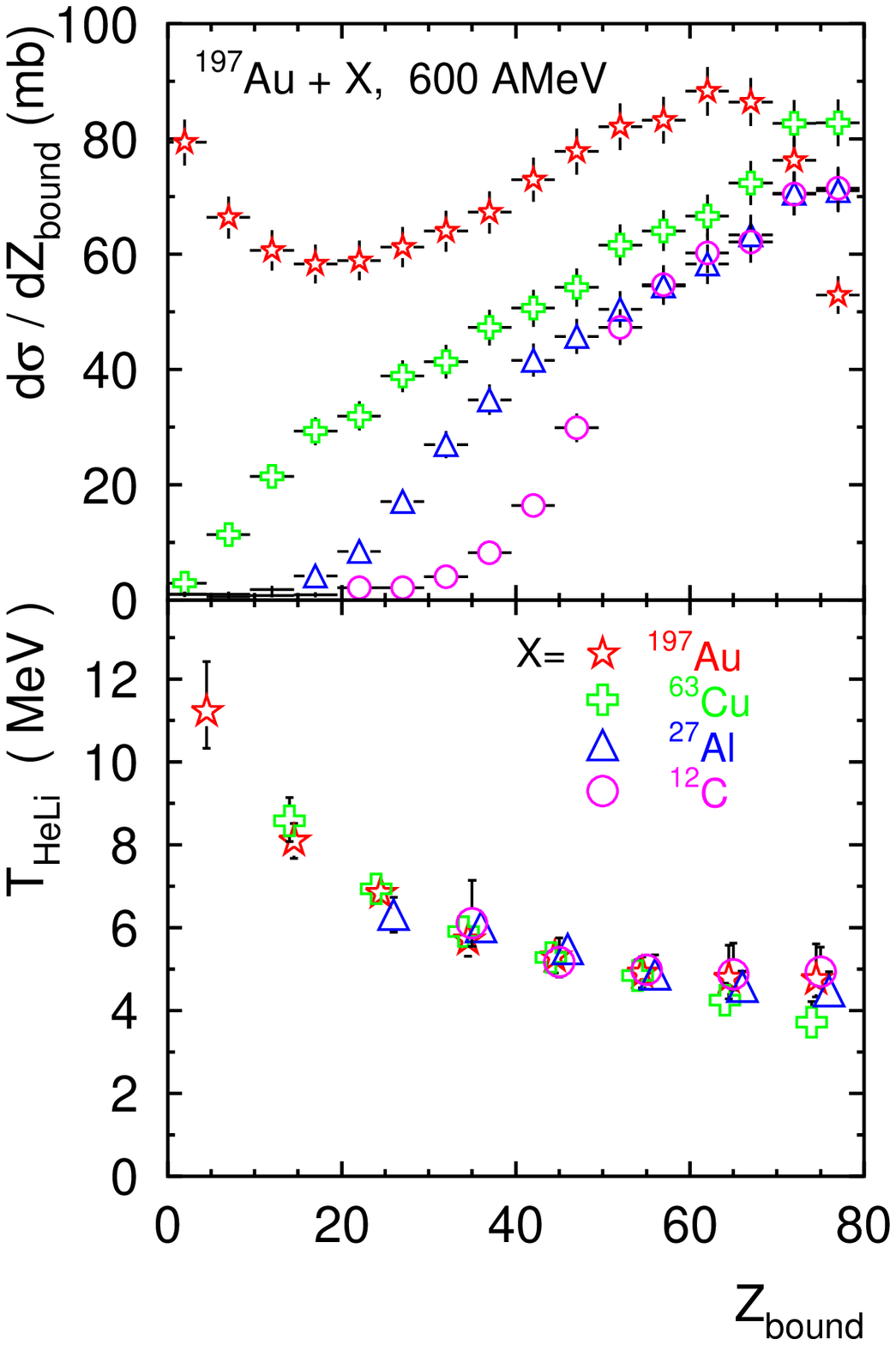}}
\caption[]{
Lower part: Isotope temperatures as a function of 
Z$_{bound}$ for Au+C,Al,Cu and Au reactions
at 600 MeV per nucleon (from ref. \cite{MOE96}). Upper part:
Differential cross section as a function of Z$_{bound}$.
}
\label{fig:14}
\end{minipage}
\end{figure}

\subsection{Caloric Curves in other Reactions}\label{sec:4_2}

The qualitative similarity of the caloric curve of water and that of
nuclei is striking. Although it is clear, that {\it this analogy should not be 
overemphasized}, it was surely this resemblance which triggered  
a widespread activity, both experimental and theoretical.
Figure \ref{fig:13} summarizes the presently available caloric curves 
measured via T$_{HeLi}$ as defined in Eq. \ref{eq:4}. 

A recent result of the EOS collaboration for $^{197}$Au+$^{12}$C reactions 
at 1000 AMeV beam energy is shown by the open circles \cite{TIN96,HAU96}.
These data nicely confirm the plateau-like behaviour
at intermediate excitation energies between 5 and 10 MeV per nucleon,
though the rise at high excitation energies is not observed in that experiment.
This is in line with a similar observation by the ALADIN
collaboration at 600 MeV per nucleon Au induced reactions on light targets
\cite{MOE96} (lower part of Fig.~\ref{fig:14}).
Though it is important to note that for the light carbon target the
cross section strongly drops for Z$_{bound}$ values below about 40
(see circles in the upper part of Fig.~\ref{fig:14}).
At small Z$_{bound}$, fluctuations in the decay as well as in the 
detection process might diminish 
the sensitivity of the event characterizing observable
(here Z$_{bound}$) to the actual initial excitation energy
for the central reactions in asymetric systems.
As a consequence, no reliable temperature values can be extracted
from the ALADIN data for Z$_{bound}$ values less than 30. 
If also for the excitation energy
the Z$_{bound}$ universality holds this means that only the 
`plateau' region can be probed by C+Au reactions.

The different reaction geometry represents a further possible source for
the deviation between Au+Au and Au+C reactions: in Au+C reactions
the participant and spectator regions have a larger overlap in
momentum space. In addition,
spectator nuclei produced in Au+Au reactions might 
be more compact compared to more rarified spectators in the most central
Au+C collisions.

A {\it preliminary} result of the INDRA collaboration 
for the Ar+Ni system at E/A=95 MeV \cite{PET96,AUG96} is indicated by 
the open triangles and squares in Fig. \ref{fig:13}.
In this reaction, the half of the projectile-like source
pointing into the beam direction has been analyzed.
While the caloric curve of this quasi-projectile
exhibits the qualitative behaviour of the 
ALADIN caloric curve, the temperature appears to be 
systematically higher by about 1-2 MeV. 
Clearly, more systematic studies are needed in order to clarify 
whether this discrepancy is for example due to the
definition of the decaying source (which in the Fermi-energy regime
is less clearly separated from the interaction region), the small size of the
system in the Ar+Ni reaction (about 32 nucleons \cite{PET96}) 
or the different neutron-to-proton contents of the source. 

\section{Emission Temperatures: The Breakdown of Equilibrium?}\label{sec:5}

\begin{figure}[htb]
\epsfysize=7.5cm
\centerline{\epsffile[80 30 500 450]{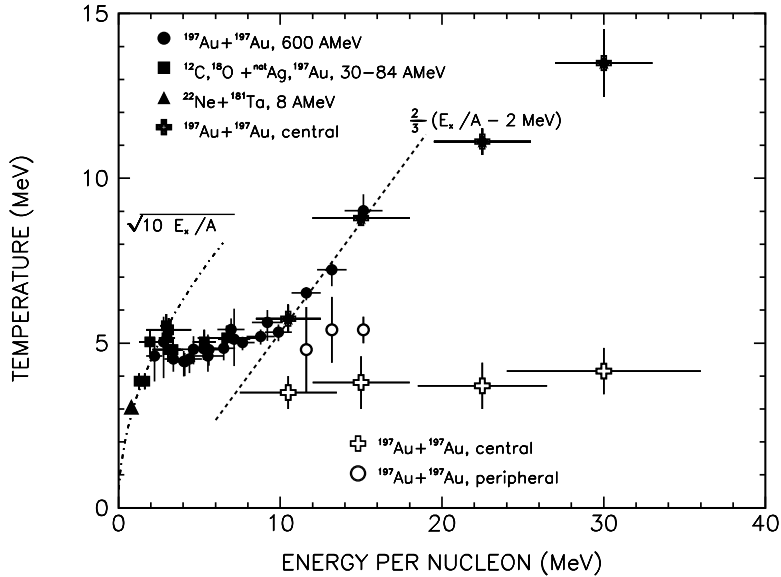}}
\caption [xxxxxx] {
Comparison of the caloric curve measured via the isotope
temperature T$_{HeLi}$ (closed symbols) with apparent emission temperatures
deduced from the relative population of states in $^5$Li
(open symbols; preliminary result from ref. \cite{SER96}). 
}
\label{fig:15}
\end{figure}

The validity of a hadronic thermometer generally rests on model assumptions. 
Only an experimental cross comparison with alternative thermometers can lend 
additional credibility to a temperature scale. For a cross calibration of the 
isotope thermometer with so called emission temperatures we, therefore, 
determined the relative population of excited states of light fragments 
in Au+Au reactions at various beam energies~\cite{SER96,RAC96}.
In figure \ref{fig:15} we compare 
the isotope temperature T$_{HeLi}$ (closed symbols) with apparent 
emission temperatures deduced from the relative population of states in $^5$Li
(open symbols).

In central collisions at beam energies between 50 and 200 MeV 
we observe a clear discrepancy between the isotope temperature 
(closed crosses) and the emission temperature (open crosses) 
which is increasing with rising beam energy. (Note that the energy scale is 
not of prime relevance for this comparison.) Besides the very low value of the 
emission temperatures of only 4 MeV, their constancy 
-- despite an increase of the beam energy by a factor of four -- is 
particularly striking.
A similar divergence of the two thermometers is seen for the
three uppermost central bins in spectator fragmentation
at 600 resp. 1000 MeV per nucleon incident beam energy.
Also there the emission temperatures (open circles) show 
a rather constant value, even though at a slightly higher level of about 5 MeV.

If the population of excited states is indeed as small
and constant as the emission temperatures suggest, sequential decays 
will only moderately disturb the isotope temperatures and, moreover, 
the relative correction will not change 
significantly with increasing excitation or beam energy. 
Surely this corroborates the isotope ratios as a robust thermometer.
But it also implies that 
- although sequential decays undoubtfully affects the difference between the 
emission and isotope temperatures --
sequential feeding alone can probably not account
for the observed discrepancy between the two thermometers \cite{TSA96}.

Lacking at the moment a quantitative explanation of this surprising 
observation, 
it might be instructive to recall a similar phenomenon during
the cosmic big-bang. Also there  
different degrees-of-freedom freeze out at
various stages of the big-bang evolution, hence signaling 
different temperatures.
On first sight such a scenario appears to be rather disappointing
since we would have to give up the idea of equilibrium between chemical
and internal degrees of freedom at freeze-out.
However, this complication may turn into an advantage in near future
since 
the various thermometer might enable us to sample the thermodynamic 
evolution of the system. 

\section{Summary and Concluding Remarks}\label{sec:7}

Disentangling collective and thermal motion in fireballs
generated in central collisions, one observes a similar caloric
behaviour as in spectator fragmentation at high excitation energies.
This concordance suggests that, despite the very different 
underlying kinematics, the multifragment decays in both reaction types 
may be governed by common physics.
Although various entrance channels lead to different caloric curves
for quasi-projectiles, 
the qualitative agreement between these curves is encouraging.
Clearly, a better knowledge of secondary processes and of the different
experimental constraints is mandatory for a further interpretation 
of the observed differences.

The shape of the caloric curve as observed by the ALADIN
collaboration is suggestive of a
first-order phase transition with a  substantial latent heat.
It also seems to exclude - on first sight - the occurrence of a second-order
phase transition which is the {\it sine qua non} for the determination of
critical-point  exponents \cite{GIL94,ELL94}.
For a finite system with $N$ constituents a transition is, on the other
hand, no longer characterized by a singular point but acquires a finite
width over a  temperature range $\Delta T_c$ around the transition
temperature $T_c$.
In case of a second order phase transition, $\Delta T_c/T_c$ is
approximately given by $\Delta T_c/T_c \sim 1/\sqrt{N}$ \cite{IMR80}. For
typical nuclear sizes of N$\approx$100, the influence of a second order
phase transition may, therefore, be perceptible at temperatures  deviating
by as much as 10\% from the critical value.
Furthermore, for a classical van der Waals gas the latent heat  increases
close to the critical point with $\sqrt{1-T/T_c}$ and reaches  already at
$T/T_c=0.95$ values larger than $k_BT_c$.
Thus, in finite systems typical features of a first-order phase transition
- like a latent heat - and signals indicating the proximity of the critical
point - like diverging moments - are not necessarily inconsistent, making
the attempt to extract critical-point exponents -- at least --  a worthwhile
and interesting venture. Time must show, whether the hope to  
apply the concept of criticality to such small systems 
like nuclei is justified.

\begin{figure}[t]
\epsfysize=7.0cm
     \centerline{\epsffile[0 20 500 500]{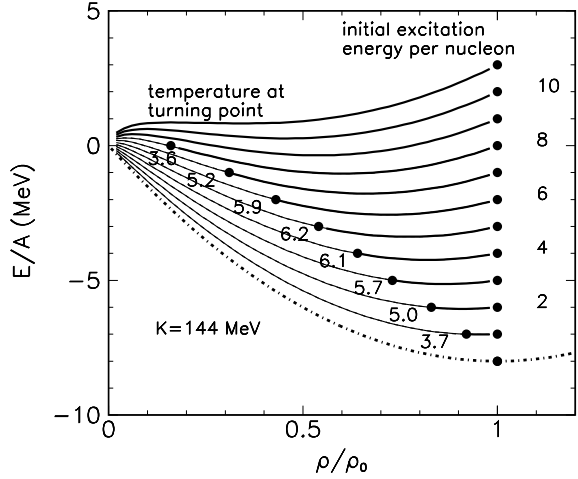}}
\caption [xxxxxx] {
Evolution of an isentropically expanding Fermi gas in the energy per nucleon
vs. density plane.
}
\label{fig:16}
\end{figure}

First hints (Fig. \ref{fig:9}) and further
experimental evidence \cite{LAC96} for the build-up of flow during 
a thermal driven decay of spectator nuclei have been found. 
This calls for a dynamical treatment of the fragmentation process.
The dynamical evolution of the heated spectator nucleus
will depend on its initial excitation energy and will, therefore,
influence or even dominate the shape of the caloric curve.
For illustration, the lines in Fig. \ref{fig:16} show
the evolution of the summed thermal and potential energy
for an isentropically expanding Fermi gas as a function of its
density; the difference with respect to the initial energy at normal
density (numbers on the right hand side)
corresponds to the energy stored (momentarily) in the collective
expansive motion.
For the low density equation-of-state of the finite
nuclear system, a parabolic density dependence

\begin{equation}
(E/A)_{T=0} = K_c/18 \cdot (1 - \rho/\rho_0)^2 - 8 MeV
\label{eq:5}
\end{equation}

\noindent
with a compressibility $K_c$ = 144 MeV was used \cite{FRI92}.
Furthermore, we ignore any dissipative processes during the expansion.

For initial excitation energies below about 8 MeV per nucleon
the system is not able to expand freely to zero density but will
stop at a finite density before it collapses again.
Since at the turning point the Coulomb barrier reaches its minimum value and
since the system spends most of its time in this region, it is probably
reasonable to assume that the fragmentation happens at this point.
Indeed, the temperatures at the turning point lie, rather independent
of the initial excitation energy, in the region of about 5 - 6 MeV.
For excitation energies beyond 8 MeV per nucleon, the system
may energetically expand to zero density prior to its freeze-out.
In such a case, the initial energy would be subdivided in a potential energy
of 8 MeV per nucleon and a collective flow energy.
In our experiment,
we do, however, observe high temperatures in the gas phase of the caloric
curve. These high values signal a freeze out at a finite density,
which is not too surprising considering the finite range
of the strong interaction and considering the fact that the
expansion velocity is still moderate.
Within a more detailed calculation \cite{PAP94,PAP96} the temperature of the
`gas branch' of the caloric curve
can be explained with a nearly constant freeze-out density of
about 1/3 of normal nuclear matter density.

While this interplay between the expansion dynamics
and the density dependent properties of a Fermi gas
might help to elucidate the gross features of the caloric curve
one has to keep in mind that we are not dealing with a homogeneous 
system. An internally consistent equation-of-state
taking into account the clusterisation \cite{PEI91},
the particle loss during the expansion \cite{FRI90,PAP94},
the viscosity of nuclear matter, and
the systematic variation of the source size in peripheral
collisions as well as the
explosive initial flow in central collisions \cite{PAP94}
is required before more definite conclusions can be drawn.
Nonetheless, if this scenario holds, the amount of radial motion
which is transferred during the expansion into collective motion,
will reflect both, the freeze-out density and the degree of
dissipation during the expansion. Particle interferometry might 
provide another, independent, determination
of the density of the system at freeze-out. Thus a combination
of all these experimental informations might
allow a closer look at the expansion dynamics.

Last but not least, a new, intriguing question resulted from the latest ALADIN 
experiment~\cite{RAC96}: Why are the emission temperatures so low 
and -- perhaps even more puzzling -- why are they constant in central collisions 
despite a variation of the beam energy by a factor of four?
While the associated small feeding contributions may be viewed as 
good news for the validity of the isotope thermometer, 
it clearly exemplifies that we are still far 
away from a detailed understanding of the thermodynamics 
of a finite, decaying nucleus.

All these open questions -- and probably many more --
will find their analogy in atomic cluster physics, cosmology and
high energy particle/nuclear physics, where the concept of
{\it phase-transition} is equally important.
In this respect, a deeper understanding of the liquid-gas phase 
transition of nuclei might also shed some light on the past history 
of our universe as well as the world around us.
Clearly, we could not provide an answer to the question
raised in the beginning. All what we can definitely say is that
nuclei do {\it not} behave just like ordinary water. 
Instead we are rewarded by the experience of a rich variety of phenomena 
which originate -- to a large extent though not exclusively -- 
from the finite size of nuclei or,
as expressed by Lucretius~\cite{LUC} when abandoning 
Empedocles' denial of void:
{\it `` nam corpora sunt et inane"}.  

\section*{Acknowledgments}
This work was supported in part by the European
Community under
contract ERBCHGE-CT92-0003 and ERBCIPD.
J.P. and M.B. acknowledge the financial support
of the Deutsche Forschungsgemeinschaft under the Contract No.
Po256/2-1 and No. Be1634/1-1, respectively.
We are especially grateful to W.G. Lynch, L.G. Moretto, 
J. Peter, E. Plagnol, and M.B. Tsang,  
for assistance in preparing this talk and for extensive discussions.


\begin{thebibliography}{99}
\bibitem{50MEV}
Except for the trigger condition (at least two particles 
in the hodoscope~\cite{RAC96}) no additional centrality cut was applied 
at E/A=50 MeV in the present stage of the analysis.
\end{thebibliography}
\end{document}